\documentclass[preprint2]{aastex62}

\usepackage{graphicx, url, times}
\usepackage{booktabs, threeparttable}
\usepackage{multirow}

\graphicspath{{pdf/}}

\newcommand{\papertitle}{The Three Hundred Project: The influence of environment on simulated galaxy properties}
\newcommand{\ahf}{\textsc{ahf}}

\newcommand{\hkpc}{{\ifmmode{h^{-1}{\rm kpc}}\else{$h^{-1}$kpc}\fi}}

\def\Mpc{\rm Mpc}
\def\Gpc{\rm Gpc}
\def\Msun{M_{\odot}}

\newcommand{\affa}{School of Physics and Astronomy, Sun Yat-sen University, 519082, Zhuhai, China}
\newcommand{\affb}{School of Physics \& Astronomy, University of Nottingham, Nottingham NG7 2RD, UK}    
\newcommand{\affc}{Departamento de F\'isica Te\'{o}rica, M\'{o}dulo 8, Facultad de Ciencias, Universidad Aut\'{o}noma de Madrid, 28049 Madrid, Spain}
\newcommand{\affd}{Centro de Investigaci\'{o}n Avanzada en F\'{\i}sica Fundamental (CIAFF), Universidad Aut\'{o}noma de Madrid, 28049 Madrid, Spain }
\newcommand{\affe}{International Centre for Radio Astronomy Research, The University of Western Australia, 35 Stirling Highway, Crawley, Western Australia 6009, Australia}
\newcommand{\afff}{University Observatory Munich, Scheinerstra{\ss}e 1, 81679 Munich, Germany}
\newcommand{\affg}{Max-Planck-Institute for Extraterrestrial Physics, Giessenbachstrasse 1, 85748 Garching, Germany}

\newcommand{\affi}{Leibniz-Institut f\"{u}r Astrophysik,
    14482 Potsdam, Germany}
\newcommand{\affj}{Department of Physics, Sapienza Universit\`{a} di Roma, p.le Aldo Moro 5, I-00185 Rome, Italy}

\received{June 20, 2018}
\revised{Sept 14, 2018}
\accepted{ }
\submitjournal{ApJ}
\shorttitle{Influence of Environment}
\shortauthors{Yang et al.}

\newlength{\figwidth}
\newlength{\resplot}
\hypersetup{ pdfauthor={Yang Wang}, pdftitle={\papertitle},
pdfsubject={pdfsub}, urlcolor=blue, }

\begin{document}

\title[Influence of environment]{\papertitle}
\correspondingauthor{Yang Wang}
\email{wangyang23@mail.sysu.edu.cn}

\author[0000-0002-1512-5653]{Yang Wang}
\affil{\affa}

\author{Frazer Pearce}
\affil{\affb}

\author{Alexander Knebe}
\affil{\affc}
\affil{\affd}
\affil{\affe}

\author{Gustavo Yepes}
\affil{\affc}
\affil{\affd}

\author{Weiguang Cui}
\affil{\affc}

\author{Chris Power}
\affil{\affe}

\author{Alexander Arth}
\affil{\afff}
\affil{\affg}


\author{Stefan Gottl\"{o}ber}
\affil{\affi}

\author{Marco De Petris}
\affil{\affj}

\author{Shaun Brown}
\affil{\affb}

\author{Longlong Feng}
\affil{\affa}


\label{firstpage}
\begin{abstract}
    The relationship between galaxy properties and environment is a widely
    discussed topic within astrophysics. Here we use galaxy samples from
    hydrodynamical re-simulations to examine this relationship. We use the
    over-density ($\delta_1$) within a $1 h^{-1}\Mpc$ sphere around a galaxy
    to evaluate its environment. Then the relations between galaxy
    properties, such as specific star formation rate(sSFR), fraction of star
    forming galaxies, $g-r$ colour and $\delta_1$ are examined within three
    galactic samples formed from galaxies within large clusters, those in the
    vicinity of large clusters and those in the field. We find tight
    environmental correlations for these galaxy properties. In brief,
    galaxies in denser environments tend to be redder and are
    more likely to be quenched. This is consistent with observations. We find that although the sSFR decreases with $\delta_1$, this is mainly because that
    galaxies with higher stellar mass reside in environment with higher overdensity.
    At fixed over-density a galaxy's colour is also independent
    of whether it lives within a cluster or within the field, but the
    relative fractions of the two samples varies dramatically with
    over-density and this drives an apparent evolution.
\end{abstract}

\keywords{
	methods: numerical -- galaxies: clusters:general -- galaxies:
	evolution
}


\section{Introduction} 
\label{sec:intro}
An important issue for the study of galaxy formation is the relationship
between a galaxy's properties and the environment it inhabits. This issue was
perhaps first studied in the pioneering work of \citet{Oemler1974} and
\citet{Dressler1980}. Their work identified and quantified a
morphology-density relationship, showing that spiral galaxies prefer to
reside in lower density environments than elliptical and S0 galaxies. Later
work investigated this relationship in more detail and explored star
formation, colour, and morphology as additional discriminants. Many of these
studies support a strong relationship between galactic properties and
environmental density \citep[e.g.][]{Kauffmann2004, Tanaka2004, Elbaz2007,
Peng2010, DeLucia2012b, Darvish2014}. Such studies indicate that galaxies
that are located in dense environments tend to be red, elliptical, and have
lower star formation rates than their low density counterparts
\citep{Lai2016}.

The observed correlation between a galaxy's properties and the environment it lives in is thought to be due to interactions with the surroundings.  Many physical processes will lead to the observed property-density relations, such as dynamical friction \citep{Chandrasekhar1943a, DeLucia2012, Contini2012b}, ram pressure stripping \citep{Gunn1972, Quilis2000, McCarthy2008}, high-speed galaxy encounters \citep[galaxy harassment;][]{Moore1996} and galaxy-galaxy mergers \citep{Mihos1994b}. However, due to the complexity of modelling the environmentally driven processes and the limitations of the observations, the relative significance of each possible environmental effect is still unclear. Selection effects and various approaches to characterizing the observed galactic density field introduce bias into the measured property - environment relationship. On the other hand, any environmental effect is a combination of many physical processes, and each galaxy will experience various environments during its lifetime \citep[see][]{DeLucia2012}, which makes any theoretical calculation extremely complex. Thus, this is an interesting multi-faceted problem from both a simulation and observational standpoint.

Several groups have investigated the influence of a galaxy's environment using semi-analytic models. For example, \cite{DeLucia2012} studied the environmental history of group and cluster galaxies. They found that the stellar mass and star formation rate (SFR) could be related to the environment a galaxy inhabited before it was accreted into its final dark matter halo. They quantified that such 'pre-processing' had an effect on $27$ -- $44$ per cent of their group/cluster galaxies, with the percentage varying for galaxies with different stellar mass. It has also been suggested that the observed difference in the fraction of passive galaxies in different clusters, as well as some other properties of galaxies, \citep[e.g.][]{Weinmann2011a, DeLucia2012b}, could be related to halo-to-halo scatter in the way the final halo was assembled (essentially a wide variation in the range of possible halo merger trees).

However, semi-analytic models do not resolve all the complex physical process involved in galaxy formation as they generally prescribe some of the relations between a galaxy and its host halo. These relations may combine the effects of many physical processes and decomposing them into analytical schema can prove difficult. On the other hand, exploring the environment-galaxy relationship with semi-analytic models can be undertaken even if the halo-galaxy relationship is manually determined so long as care is taken not to simply recover an input relationship. To go beyond this cosmological hydrodynamical simulations are commonly used. Hydrodynamical simulations, which include baryonic processes operating under a self-consistent gravitational framework, directly encode many of the physical processes that may impinge upon galaxy formation and evolution. Until recently, there haven't been many studies investigating the galaxy property - environment relationship using hydrodynamical simulations. \cite{Gabor2012,Gabor2015} reproduced the environmental dependence of galaxy colour and quenching within a hydrodynamical simulation, finding a larger fraction of red galaxies within denser environments. Their results indicate that satellite galaxies are affected by environmental quenching, while the quenching process operating for central galaxies is largely driven by their own stellar mass. Later work, \cite{Vogelsberger2014}, investigated the same processes as \cite{Gabor2012,Gabor2015} within the larger \textsc{Illustris} simulation. Both these studies agreed with the observational results from SDSS and zCosmos surveys \citep{Peng2010} quite well. Later \cite{Rafieferantsoa2015} investigated the relationship between $\rm H\,\textsc{\lowercase{I}}$ content and local overdensity in a large simulation. This work indicated that the median $\rm H\,\textsc{\lowercase{I}}$ and specific star formation rate (sSFR) drops in denser environments, whether or not quenched galaxies are taken into account. This result is also consistent with observational studies \citep{Fabello2012}. Lotz et al.(2018 in prep), utilizing the \textsc{Magneticum} simulation, found that star forming galaxies are quenched during their first passage. Further they find that quenching is impeded at high stellar mass, suggesting that the mass of such galaxies effectively shields them from ram-pressure stripping. While hydrodynamical simulations, such as those mentioned above, are considered a vital tool in aiding and interpreting astronomical observations of galaxy clusters \citep{Borgani2011}, the detailed effect of baryonic physics remains unclear, especially on small scales \citep[e.g., see][for a review of baryonic effects ]{Cui2016, 2017tmc..book....7C}. Furthermore, the use of different simulation codes and techniques and the inclusion (or not) of different physical processes adds additional uncertainty. The nIFTy galaxy cluster comparison project \citep{Sembolini2016a,Sembolini2016, Elahi2016, Cui2016,Arthur2016} compared a dozen common simulation codes, by simulating one identical galaxy
cluster to investigate their differences. They found that although the
dark-matter-only runs gave quite good agreement between different simulations
\citep{Sembolini2016}, the hydrodynamical runs showed a large discrepancy,
especially when different baryonic models are included \citep{Cui2016}. Since
one of the main issues with the nIFTy comparison was the potential for
cluster-to-cluster scatter, the {\sc Three Hundred Project} aimed to increase
the galaxy cluster sample to 324 large clusters. It focuses on the
statistical results for both hydrodynamical simulations and semi-analytical
galaxy formation models of the clusters \citep{Cui2018}. As one work in this
project, we focus on quantifying any environmental effect on the galaxies
within our hydrodynamical simulations. Two codes, {\sc Gadget-X} and {\sc
Gadget-MUSIC}, are employed for running the re-simulations. As well as the
324 cluster re-simulations, we also simulate four large field regions
specifically targeted so as not to include any significant cluster. This will
allow us to explore the possible influence from large scale structures. We
intend to study the relationship between galaxy properties and their
large-scale environment.

The remainder of this paper is organized as follows: in section \ref{sec:data} we describe the simulations we have used and the definition of the extracted galaxy properties and environmental density. In section \ref{sec:results} we present the results of our analysis; \ref{sec:results:sfr} discusses the relation between SFR and environmental over-density. \ref{sec:results:c} shows the environmental dependence of a galaxy's colour and magnitude. Finally we draw conclusions in section \ref{sec:summary}.

\section{Simulation Data} \label{sec:data}
This paper utilizes the simulation dataset provided by the {\sc Three Hundred
Project}. This consists of 324 re-simulated clusters and 4 field regions
extracted from within the MultiDark Planck simulation, MDPL2,
\citep{Klypin2016}. The MDPL2 simulation has cosmological parameters
of $\Omega_M=0.307, \Omega_B=0.048, \Omega_{\Lambda}=0.693, h=0.678, \sigma_8=0.823$.
All the clusters and fields have been simulated using the
full-physics hydrodynamical codes {\sc Gadget-X} and {\sc Gadget-MUSIC},
which are updated versions of {\sc Gadget2} \citep{Springel2005a}.
In the re-simulation region, the mass of a dark matter particle is $12.7\times10^8h^{-1}\Msun$ and the mass of a gas particle is $2.36\times10^8h^{-1}\Msun$. 
The mass of a star particle varies from $3.60\times10^7h^{-1}\Msun$ to $1.65\times10^8h^{-1}\Msun$ with $99\%$ of the star particles being less massive than $4.60\times10^7h^{-1}\Msun$.
Each cluster re-simulation consists of a spherical region of
radius $15h^{-1}\Mpc$ at $z=0$ centred on one of the 324 largest objects from
within the host MDPL2 box, which is $1h^{-1}\Gpc$ on a side. The host halos of our galaxies range in mass from
$2.54\times10^{10}h^{-1}\Msun$ to $2.63\times{10}^{15}h^{-1}\Msun$.
The largest halo within each of the 324 cluster re-simulations varies from $8.15\times10^{14}h^{-1}\Msun$ to $2.63\times{10}^{15}h^{-1}\Msun$. The field
regions are low density volumes of radius $43h^{-1}\Mpc$ selected so as not
to include any halos larger than $4 \times 10^{13}h^{-1}\Msun$. Although not
technically selected to be voids they are all under-dense relative to the
cosmic mean. 
Detailed descriptions of the 324 clusters and the simulation codes can be found in \cite{Cui2018}. 
We also refer interested readers to \cite{Beck2016} for details
of {\sc Gadget-X} and \cite{Sembolini2013} for details of {\sc Gadget-MUSIC}. In
summary, {\sc Gadget-X} uses an improved SPH scheme \citep{Beck2016}, while {\sc Gadget-MUSIC} uses a classic SPH scheme;
{\sc Gadget-X} includes a full physics baryon model including AGN feedback, while {\sc
Gadget-MUSIC} does not take massive black hole growth or AGN feedback into
account; in addition the models for stellar feedback are different . In this paper, we concentrate on the results of {\sc Gadget-X}
because, as we show below, this approach better reproduces the observed
distribution of galaxy mass as a function of star formation rate.

To define a galaxy, we first use \ahf\ \citep{Knollmann2009} to find halos
and sub-halos within the simulations. Then a group of star and any associated
gas particles inside a sub-halo is defined as a galaxy. For central galaxies
which do not belong to any sub-halos, we follow \cite{Furlong2015} by
defining the group of all star and gas particles within $30 \hkpc$ from the
halo center as the central galaxy. While in principle a galaxy could contain
as little as one star particle such poorly resolved galaxies will bring large
uncertainties to any measurement of their properties. Thus, we constrain our
galaxy samples by imposing a lower mass limit of $10^{9.5}M_{\odot}/h$, which
roughly corresponds to $80$ stars. We also excluded any galaxies whose host
halo was contaminated by boundary particles. Boundary particles are low
resolution particles used to surround the re-simulated region in order to
preserve the influence of large scale structure from the entire MDPL2 volume.
Since we need to calculate each galaxy's environment, we further abandon any
galaxies lying too close to the boundary to avoid sampling any regions not populated by galaxies
or contaminated with boundary particles.
In another words, we only selected galaxies within  
$10h^{-1}\Mpc$ of the center of cluster re-simulations.
For field re-simulations, the corresponding distance is $38h^{-1}\Mpc$.

The largest galaxy in the cluster re-simulations has a stellar mass
(including its associated intra-cluster light (ICL)) of $10^{13}h^{-1}\Msun$.
In the field regions this value is only $\sim 10^{11}h^{-1}\Msun$. The
number of galaxies in each sample group, above the indicated stellar mass,
are represented in Table~\ref{tab:galaxy}. The number of galaxies above a fixed stellar mass within the {\sc Gadget-MUSIC} simulations is almost twice that of {\sc Gadget-X}. This is because, in the absence of AGN feedback, star formation within {\sc Gadget-MUSIC} is more efficient. Essentially, as we see below, the galaxies within the {\sc Gadget-MUSIC} models are more massive than their counterpart in the {\sc Gadget-X} models.

\begin{table}
	\begin{center}
		\caption{Number of galaxies at $z=0$ with stellar mass above $10^{9.5}h^{-1}M_{\odot}$ within each sample class as indicated for {\sc Gadget-X} (left column) and {\sc Gadget-MUSIC} (right column).}
		\label{tab:galaxy}
		\begin{tabular}{c c c }
			\toprule
                & {\sc Gadget-X} & {\sc Gadget-MUSIC} \\
			\midrule
            Cluster & 96470 & 133105\\ 
            Vicinity & 111662 & 186321\\ 
            Field & 7365 & 13724\\ 
			\bottomrule
		\end{tabular}
	\end{center}
\end{table}

\begin{figure*}
	\includegraphics[width=1\linewidth]{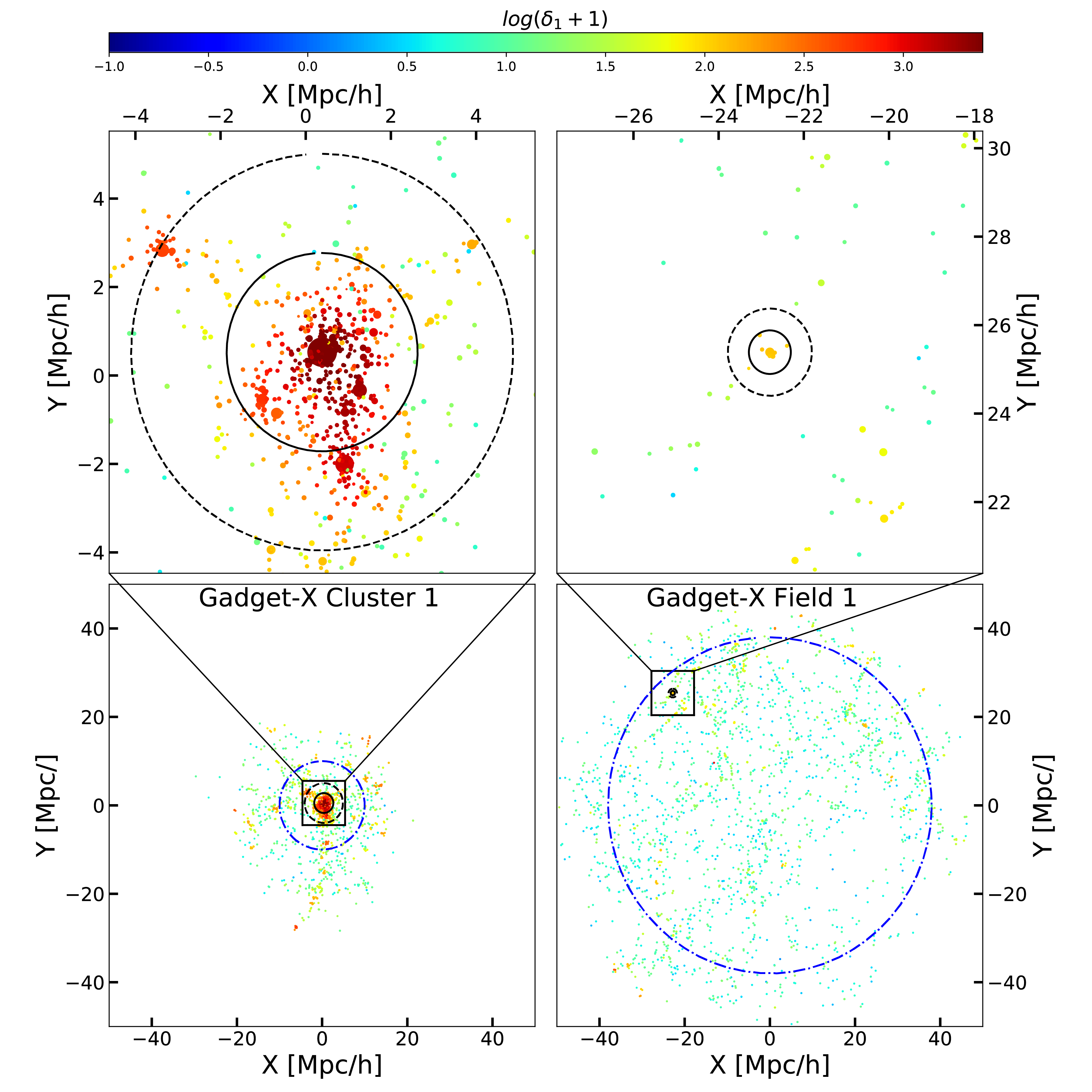}
	\caption{
        The distribution of galaxies within Cluster 1 and Field 1 at $z=0$, taken from the {\sc Gadget-X} simulation. Each circle represents a galaxy with the circle's size proportional to its stellar mass . The environmental over-density within $1h^{-1}\Mpc$($\delta_1$) is indicated by the galaxy colour. Only galaxies with stellar mass above $10^{9.5}h^{-1}\Mpc$ are shown. The bottom panels show an overview of the full region being re-simulated and to aid comparison have the same linear size. The top panels show a zoom into $10h^{-1}\Mpc$ sided cubes centered on the most massive galaxy. The solid and dashed black circles indicate $R_{200}$ and twice $R_{200}$, the radius and double the radius of the most massive dark matter halo in each re-simulation. In both cases galaxies outside the blue dashed circle are excluded from the sample, as described in the text.
	}
	\label{fig:img}
\end{figure*}

Galaxies from our simulations are split into three categories based on their physical location. Galaxies located within twice $R_{200}$ of the cluster centre are regarded as `cluster galaxies'. Galaxies outside this radius in the cluster re-simulations are regarded as `vicinity galaxies'. Galaxies in the field re-simulations contribute to the `field galaxy' sample. 
The re-simulation volume for consideration in the following text is always set to be a co-moving sphere with a radius of $10h^{-1}\Mpc$ for the cluster re-simulations or $38h^{-1}\Mpc$ for the field re-simulations.
The centre of these outer spheres is taken to be the center of the
re-simulation region at $z=0$ and fixed for all time. The cluster center is
found by AHF and moves with time. Hence the volume extracted as the 'cluster region' moves in both location and size as the simulation evolves, although it remains well within the outer boundary at all times. The cluster 'vicinity' region is all the volume that is outside the cluster region boundary and within the outer boundary.

To give a visual impression of our re-simulations, images of the largest
cluster and one of the field re-simulations at $z=0$ are shown in
Figure~\ref{fig:img}. Each circle represents a galaxy (with mass larger than
$10^{9.5}h^{-1}M_\odot$) with the size of the circle proportional to the
galactic stellar mass. The colour represents the over-density $\delta_1$ of
its local environment, in this case within $1h^{-1}\Mpc$(see detailed definition of $\delta_1$ in Section \ref{sec:data:env}). The bottom panels
give an overview of the full re-simulated region. The top panels show a zoom
into a $10(\Mpc/h)^3$ cube centered on the most massive galaxy. The blue
dashed circle indicates the volume within which we could recover an accurate
environmental over-density for each galaxy. Galaxies outside the circle are
excluded from our analysis. The radius of the blue circle is $10h^{-1}\Mpc$ for cluster re-simulations and $38h^{-1}\Mpc$ for field re-simulations. 
The two black circles indicate $R_{200}$ (solid)
and $2R_{200}$ (dashed) of the most massive halo in each simulation. Their position and size will change through time.
This figure also illustrates how our three galaxy classes are defined: in cluster re-simulations, galaxies within the black dashed circles are classified as `cluster galaxies', and galaxies between the black dashed circle and the blue dash-dotted circle are classified as `vicinity galaxies'. In field re-simulations, all galaxies within the blue circle form the `field galaxies' sample. Note that the sizes of all the regions are given in co-moving units. 

\subsection{Measurement of Galaxy Properties}
\label{sec:data:gprop}
We study the environment's effect on a galaxy's SFR, luminosity
(in the $r$ and $g$ bands), and colour ($M_g- M_r$).
How we define and calculate each of these properties is given below:

\begin{itemize}
     \item {\sc SFR} and {\sc sSFR}: We define the SFR of a galaxy as the sum of the SFRs of all gas particles belonging to it. In the simulation, each gas particle has its own star formation rate which is determined by the star formation model within the simulation \citep{Springel2005b}. At each step in the simulation, some fraction of the mass of a gas particle is converted into a new star particle according to its SFR. The {\sc sSFR}(specific star formation rate) is defined as usual, $sSFR = SFR/M_*$.

    \item {\sc Luminosity} and {\sc Magnitude}: The luminosity in any defined
    spectral band is calculated by applying the stellar population synthesis
    code {\sc STARDUST} \cite[see][and references therein for more
    details]{Devriendt1999}. This code computes the spectral energy
    distribution from the far-UV to the radio. The stellar contribution to
    the total flux is calculated assuming a Kennicutt initial mass function
    \citep{Kennicutt1998d}. Absolute {\sc magnitudes} are readily calculated
    from the luminosity.
    Note dust obscuration is not taken into account. The {\sc Three Hundred Project} gives the luminosities and magnitudes of galaxies in several bands. For this work we only use those derived from imposing Sloan Digital Sky Survey(SDSS) $g$ band and $r$ band filters.

\end{itemize}

\subsection{Measurement of a galaxy's environment} \label{sec:data:env}
Observationally there are three methods commonly used to characterize a galaxy's environment:
\begin{itemize}
     \item Count the number of neighboring galaxies within a projected ring around each galaxy \citep[e.g.][]{Wilman2010a}.
     \item Count the number of neighboring galaxies within a fixed volume around each galaxy \citep[e.g.][]{Kauffmann2004, Gallazzi2009, Gruetzbauch2011c}.
     \item Find the projected distance $r$ to the $n\rm{th}$ nearest neighbour galaxy, with $n$ in the range $3$--$10$. Then the projected surface density is calculated via $\Sigma_n = n/\pi r^2$ \citep[e.g.][]{Dressler1980,Hashimoto1998,Tanaka2004,Capak2007a}.
\end{itemize}
Further details and a comparison between various environmental estimators are given in \cite{Muldrew2012}. As the above methods show, observationally the traditional parameter defining the environmental density is the local galaxy number density. However, since galaxies form a biased tracer of the density field \citep{Mo1996}, we might expect to find a more intrinsic environmental dependence if we utilise the underlying dark matter density field. From observations, this underlying matter density field could be reconstructed from combining the galaxy mass density with a bias factor $b$. Galaxy mass density comes from weighting each galaxy with their mass in the formula of galaxy number density. However, the bias can be complicated depending on the scale, redshift and type of galaxies included \citep{Kovavc2010}. On the other hand, we can directly measure the actual density (smoothed on some scale) within a simulation.

In this work we define $\delta_1$, the local over-density of all matter (including stars, gas and dark matter) compared to the mean density of the universe, within a sphere of radius $1h^{-1}\Mpc$ centred on the galaxy to quantify the local environment. $1h^{-1}\Mpc$ is a characteristic radius of dark matter halos with mass around $10^{13}$ -- $10^{14} h^{-1}\Msun$. It is also a commonly used distance in the second method above for defining the environmental galaxy number density. This is the region that is thought to be tightly connected with a galaxy \citep{Kauffmann2004}. We have also measured the environment using radii of $2h^{-1}\Mpc$ and $3h^{-1}\Mpc$ to form $\delta_2$ and $\delta_3$. Such distances are widely used as the smoothing scale when calculating structural features such as nodes, filaments, sheets and voids \cite[e.g.][]{Hahn2007,Zhang2009}. We found that $\delta_1$ is linearly related to $\delta_2$ and $\delta_3$. The dependence of a galaxy's properties on $\delta_2$ and $\delta_3$ has almost the same shape as those found for $\delta_1$. Thus, in what follows we only show the results for $\delta_1$.

\begin{figure} \includegraphics[width=1\linewidth]{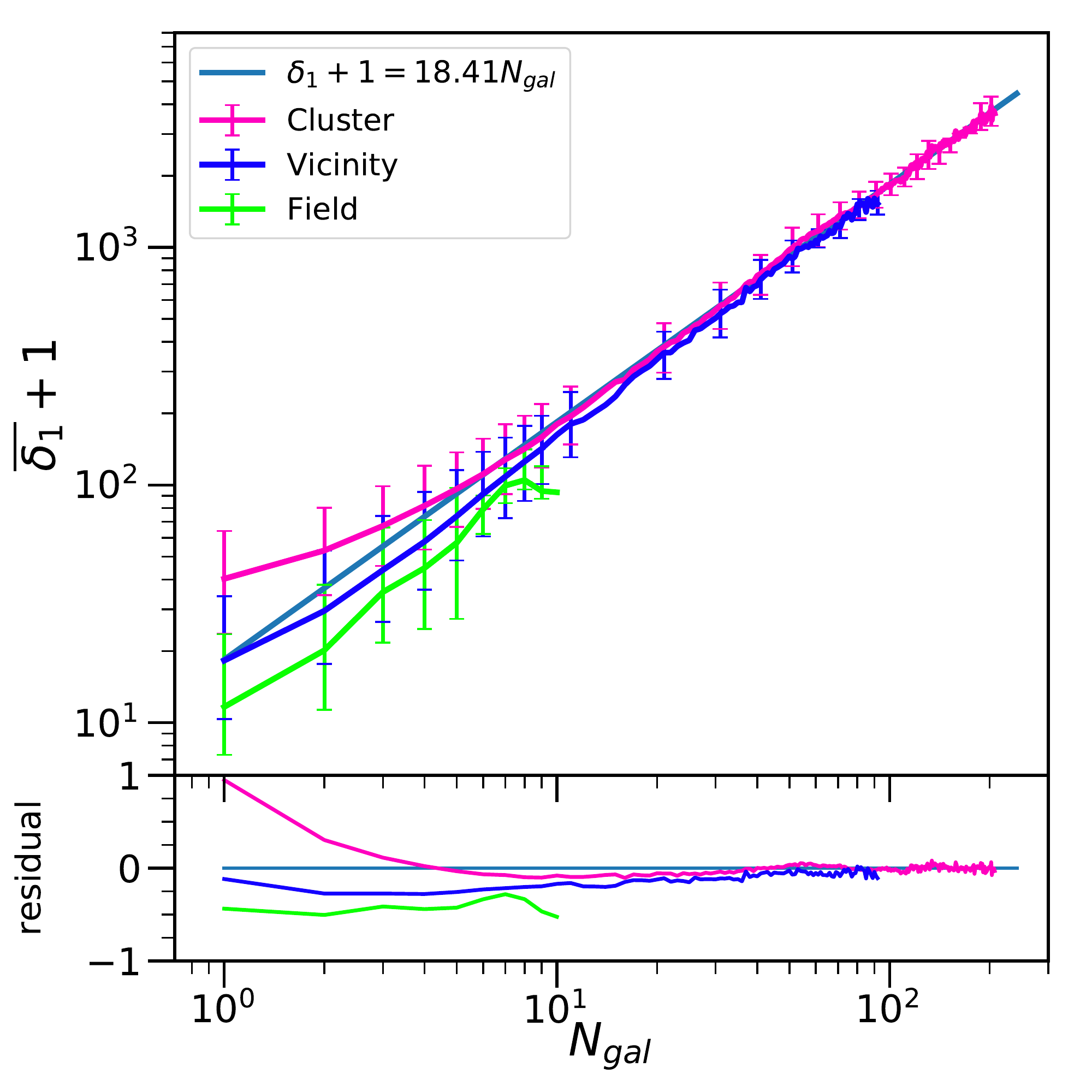}
    \caption{ 
      The median matter over-density $\overline{\delta}$ as a function of galaxy number $N_{gal}$ within $1h^{-1}\Mpc$ (top panel). The error bars show the $16^{th}$ and $84^{th}$ percentiles. The cyan line without error bars is a linear fit of $\delta$ directly proportional to $N_{gal}$ scaled to the cluster galaxy sample. In the lower panel the residuals from the linear fit are shown. The residual is calculated as $(\delta_1-\delta_{fit})/(\delta_{fit}+1)$.
    } 
    \label{fig:ngal} 
\end{figure}

To probe the relationship between the intrinsic matter over-density, $\delta_1$, and the more observationally motivated galaxy number, $N_{gal}$ we present Figure~\ref{fig:ngal} here. It shows the median over-density of $1h^{-1}\Mpc$ spheres as a function of the galaxy number in the same region. For each galaxy, we both measure $\delta_1$ and count the number of galaxies with mass above $10^{9.5}M_\odot/h$ contained in the same volume. As Figure~\ref{fig:ngal} shows, once a handful of galaxies are present, the galaxy number density is linearly related to the matter over-density. The curves of the three sample groups converge as the number of galaxies contained in the volume increases. Divergences appear when $N_{gal}$ is less than $10$. For a region with low galaxy number density, its corresponding over-density tends to be higher if it belongs to the cluster sample compared to those which belong to the vicinity and field samples. 
Within our cluster galaxy sample essentially no low over-density environment exists. This skews the distribution of galaxy number because although sometimes cluster galaxies live in relative isolation this environment is not under-dense. Additionally, within this environment, as well as the matter density potentially being increased by the presence of a large dark matter halo, gas is readily ram-pressure stripped. Thus at the edge of the cluster less gas is available to make stars and consequently fewer (and smaller) galaxies might be formed. Primarily though, out to twice $R_{200}$ matter over-densities much below 30 times the cosmic mean simply don't occur, and this distorts the distribution at low galaxy number.

As Figure~\ref{fig:ngal} illustrates, for any single galaxy there is quite a bit of scatter in the $\delta_1$ to $N_{gal}$ relationship. Generally though, these two quantities are clearly related and we can use the intrinsic quantity $\delta_1$ interchangeably with the observational measure $N_{gal}$ in what follows. In practice, any observational measure of $N_{gal}$ would use a different mass limit for the included galaxies and so the correspondence would need to be recalculated for this value. In this case though it is $\delta_1$ that is the fixed quantity, with $N_{gal}$ changing.

\section{Results} \label{sec:results}

\subsection{Specific Star Formation Rate Vs. Environmental Over-density} \label{sec:results:sfr}

\begin{figure}
    \includegraphics[width=1\linewidth]{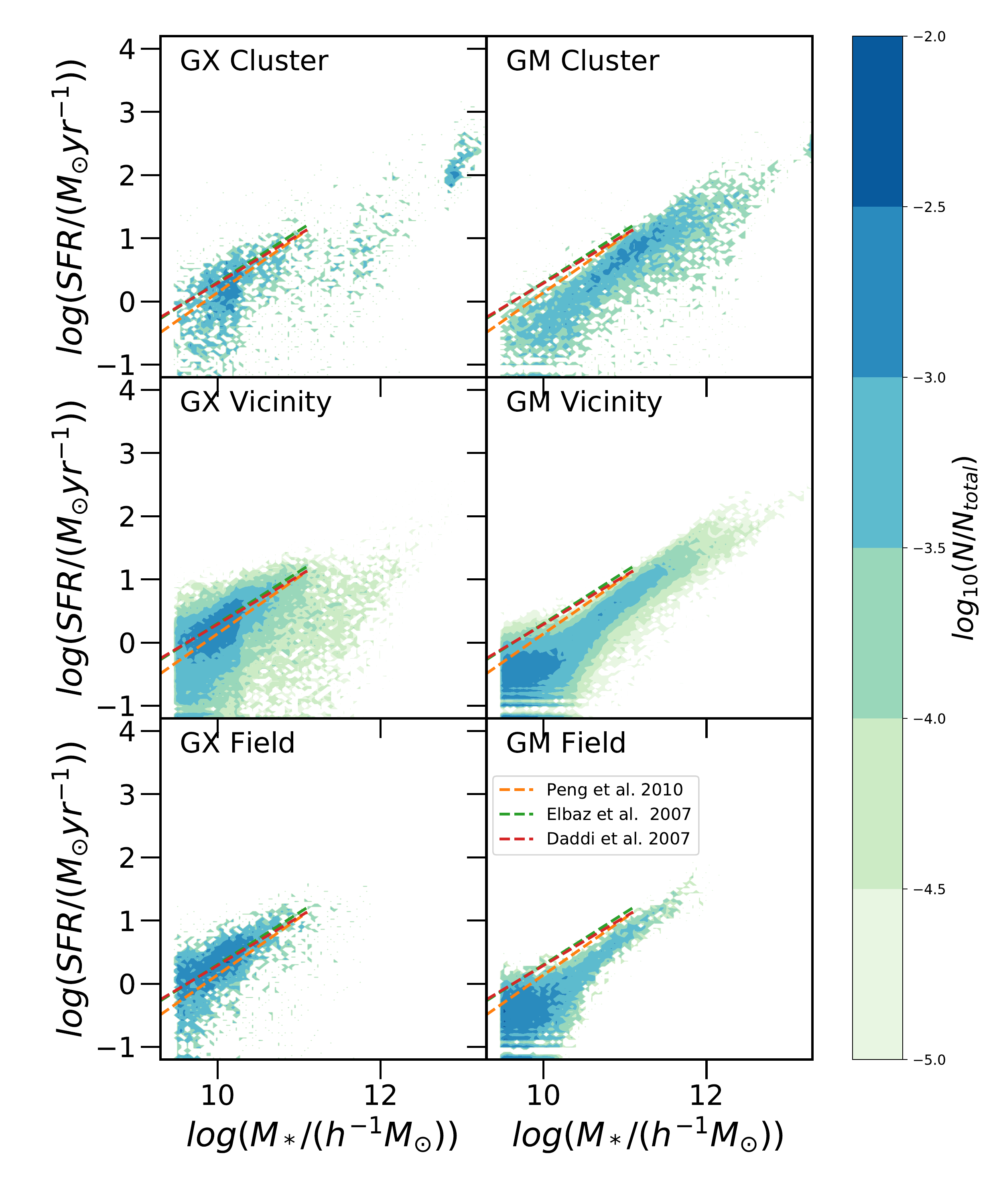}
    \caption{
        The star formation rate of galaxies as a function of galaxy stellar mass at $z=0$. The left column shows results for the {\sc Gadget-X} simulations, the right column shows results for the {\sc Gadget-MUSIC} simulations. The rows show our three environments, as indicated in the panel. The red, green and black dashed lines indicate observational fits to the SFR-$M_*$ relation taken from \protect\cite{Peng2010}, \protect\cite{Elbaz2007} and \protect\cite{Daddi2007}. According to the references, these fitting lines are valid for a rough range of $10^7\Msun<M_*<10^{11}\Msun$. The colour shading denotes galaxy number density within the sample, as shown via the colour bar on the right.
            }
    \label{fig:sfr_m}
\end{figure}
Although in this work we are principally interested in the relationship
between a galaxy's properties and the environment it lives in, we first check
that a fundamental observed relation, in this case the star formation rate to
stellar mass ($SFR-M_*$) relation, is recovered. In Figure~\ref{fig:sfr_m},
we show the $SFR-M_*$ relation at $z=0$ for our three galaxy samples, with cluster
galaxies on the top row, vicinity galaxies in the middle and field galaxies
at the bottom. The left column of panels are for galaxies within the {\sc
Gadget-X} models, the right column for galaxies from the {\sc Gadget-MUSIC}
simulations. The shading indicates galaxy density as given by the colour bar.
We overplot best fit lines for $SFR(M_*)$ derived from \cite{Daddi2007},
\cite{Elbaz2007} and \cite{Peng2010}. As Figure~\ref{fig:sfr_m} shows, the
{\sc Gadget-X} runs reproduce galaxies with a $SFR-M*$ relation in good
agreement with observations, within the stellar mass range of
$10^{9.5}$ to $10^{11}h^{-1}\Msun$. For galaxies more massive than $10^{11}
h^{-1}\Msun$ in {\sc Gadget-X} clusters, the majority are below the
observational fits. The SFR in the high mass range is supposed to be suppressed by AGN feedback.
A similar suppression may be present in the observations \cite[e.g., see Figure 17 of][]{Brinchmann2004}.

The {\sc Gadget-X} field galaxies possibly show a slightly higher SFR than
the fits but this may just reflect an apparent absence of galaxies with low
star formation rates within this sample. Note that the main sequence lines
from observations are based on samples which are mixtures of galaxies from
all our samples, while cluster and vicinity galaxies dominate our total
sample. In contrast to the reasonable looking {\sc Gadget-X} galaxies, the
{\sc Gadget-MUSIC} galaxies have an obviously lower SFR at a given mass than
the observed relations. The reason for this low SFR is somewhat
counter-intuitive in that within the {\sc Gadget-MUSIC} simulations {\it too
many} stars have been formed. However, for any given galaxy the measured SFR
is always low, even at higher redshift. The contradiction is only apparent:
for a matched object between the two simulations the SFR is higher in the
{\sc Gadget-MUSIC} simulation than in the {\sc Gadget-X} simulation, so the
matched galaxy acquires a larger stellar mass. This moves it to the right in
Figure~\ref{fig:sfr_m}, giving it an apparently low SFR {\it for a galaxy of
this mass}. Essentially, Figure~\ref{fig:sfr_m} demonstrates that the {\sc
Gadget-MUSIC} simulations produce overly massive galaxies on all scales that
are significantly more massive than those observed. Because of this for the
remainder of this paper we concentrate our results on the more physically
reasonable galaxies found within the {\sc Gadget-X} simulations.

\begin{figure}
    \includegraphics[width=1\linewidth]{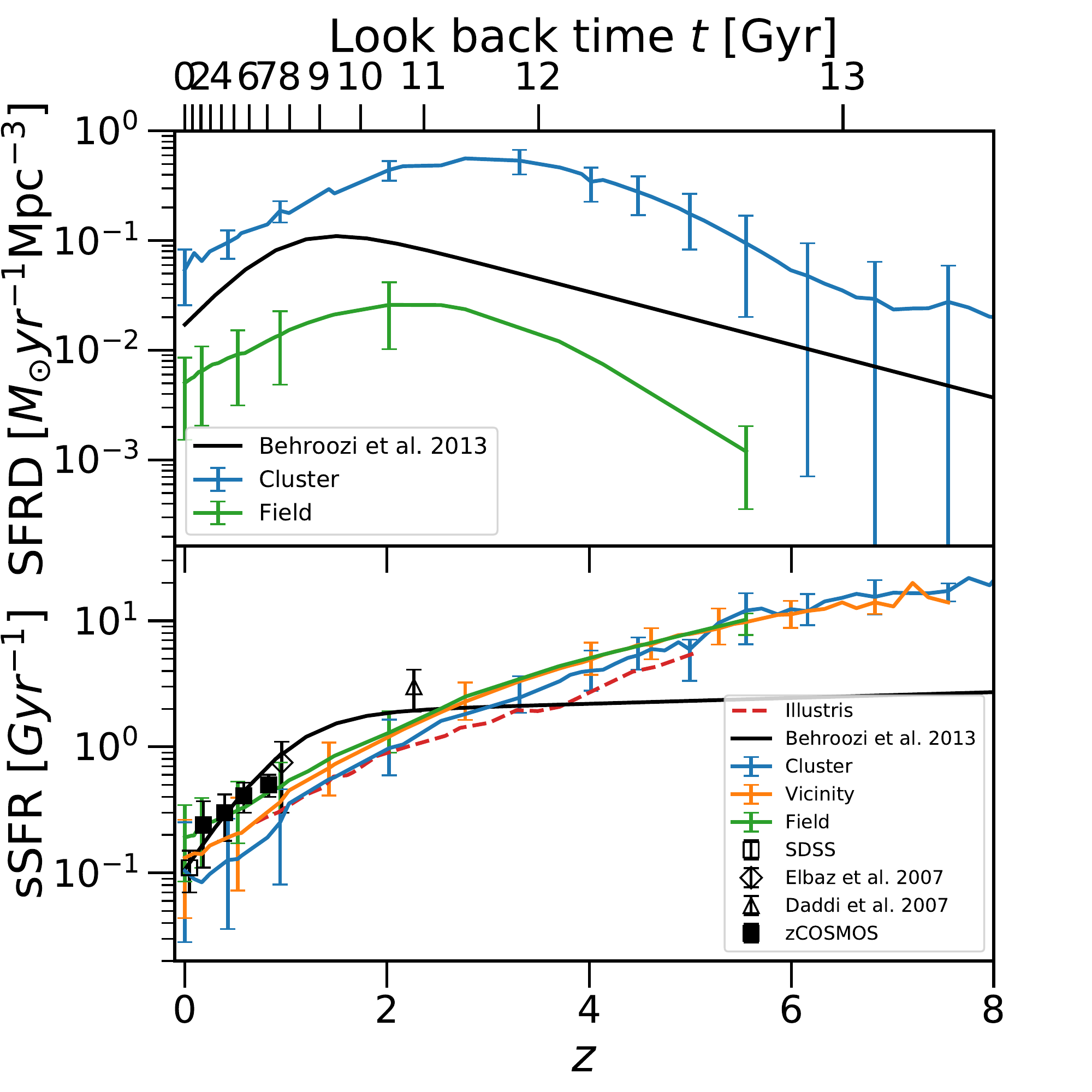}
    \caption{
        The evolution of the cosmic star formation rate density (SFRD, top panel) and median specific star formation rate (bottom panel).  Above $z=0$ the geometry of the region defining the cluster sample is not simple and so we combine the cluster and vicinity galaxy samples together. The bottom panel shows the evolution of the median specific star formation rate of galaxies with $0.8 \times 10^{10}\Msun/h < M_* < 1.2 \times 10^{10}\Msun/h $. The error bars indicate the $16^{th}$ and $84^{th}$ percentile in each redshift bin. Results from different galaxy samples are distinguished by colors. The symbols with error bars are observational data derived from figure 3 of \protect\cite{Peng2010}. In both panels, fits to observation data from \protect\cite{Behroozi2013} are plotted as black lines. The sSFR history from the {\sc Illustris} simulation \protect\citep{Sparre2015} is plotted as a red dashed line in the bottom panel.
        }
    \label{fig:sfr_z}
\end{figure}

As a further check we show the evolution of the cosmic star formation rate density (SFRD) and median specific star formation rate in Figure~\ref{fig:sfr_z}. In the top panel, the star formation rate density in the field (green line) and cluster re-simulations (blue line) span the observational fit of \cite{Behroozi2013} (black line). This is to be expected as the cluster re-simulation region by design contains a large cluster and many galaxies while the field region is devoid of any large clusters and contains fewer galaxies than average per unit volume. We have combined the cluster and vicinity galaxy samples into the same blue line as the spatial split between the two makes it complicated to determine the exact volume. To construct this volume at higher redshift we always choose a sphere of co-moving radius $10h^{-1}\Mpc$ centred on the middle of the re-simulation volume. In common with the observations, both the cluster and field SFRD curves rise to a peak at intermediate redshift and then the SFRD falls by around an order of magnitude by $z=0$.

The evolution of the median sSFR for each of our samples is given in the lower panel of Figure~\ref{fig:sfr_z}.
At $z=0$, galaxies in the field have the highest median sSFR, and cluster galaxies have the lowest median sSFR.
The difference among galaxy classes is distinguishable but with large error bars. The difference becomes smaller at higher redshifts. The three median sSFR lines finally converge at about $z\ge5$. 
The simulated sSFR follows the observational results given by the black symbols at very low redshift, then keeps below the observational symbols from $z\sim0.1$ to $z\sim2.5$ before exceeding the fitting line at high redshifts.
The line for cluster galaxies is not within the error bars of black observational data points, but 
this is reasonable because the observations take all classes of galaxies into account.
In common with other hydrodynamical simulations e.g. {\sc EAGLE} \citep{Furlong2015}, {\sc Illustris} \citep{Sparre2015} and semi-analytic models, \citep{Henriques2015}, we do not recover the double-power law fit indicated by the fit of \cite{Behroozi2013} \citep[see figure 5 in ][]{Davidzon2017}. This lack of any clear break in the median sSFR either indicates that the observational sample is as yet incomplete or that the models are too efficient at forming stars at early times\citep{Asquith2018}.

As we have also shown above in Figure~\ref{fig:sfr_m}, many previous works have indicated a tight relationship between SFR and $M_*$ with $M_*$ expected to be $\propto \rm{SFR}$, from both theoretical and observational standpoints \citep[e.g.][]{Dave2008, Wang2013, Yates2012}.  We can therefore emphasize any deviation from this relationship by calculating the specific star formation rate $sSFR=SFR_{gal}/M_*$, which removes the direct element of the mass dependence.

\begin{figure*}
    \includegraphics[width=1\linewidth]{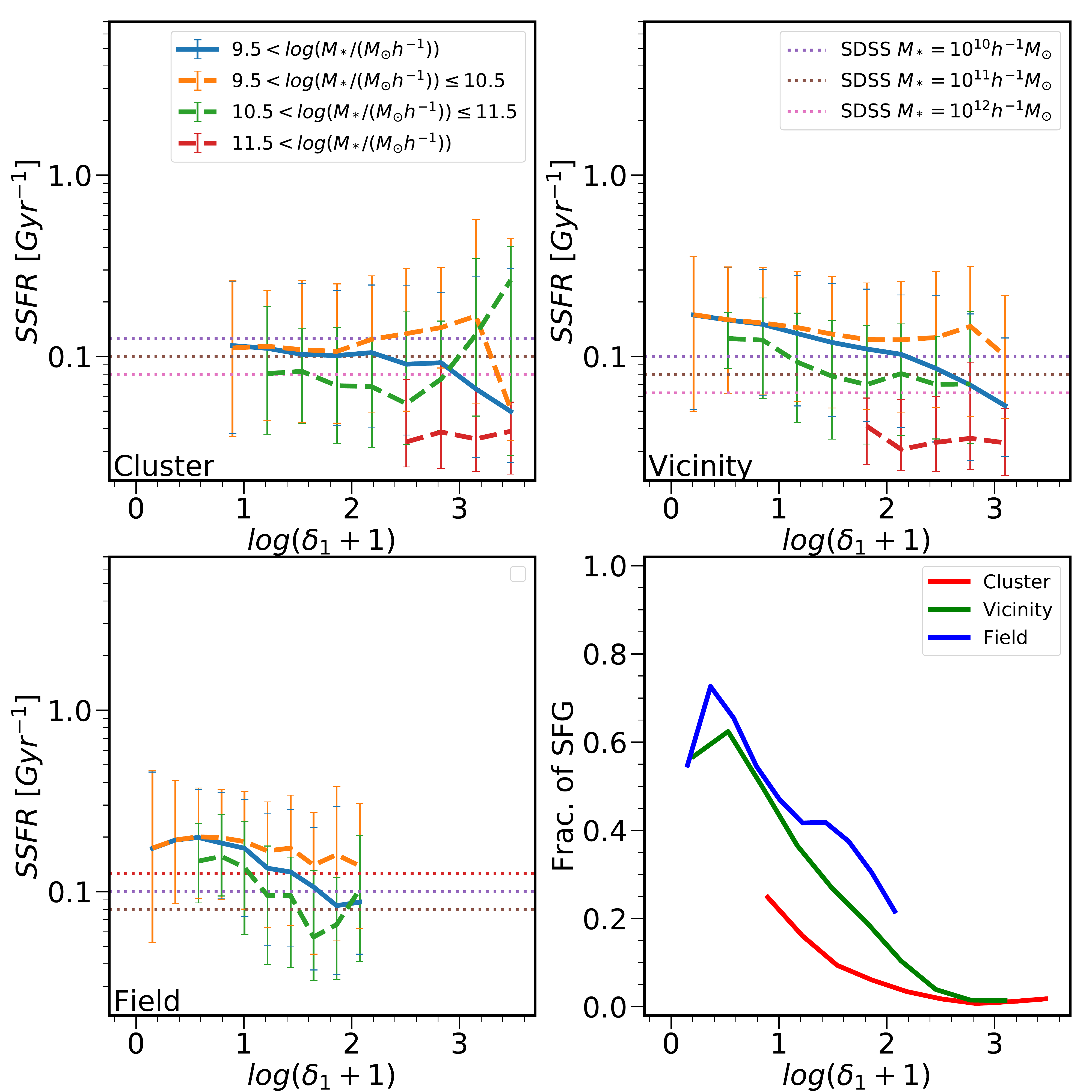}
    \caption{
        The median specific star formation rate(sSFR) of star forming galaxies (SFG) for all galaxies above $10^{9.5}h^{-1}\Msun$ (blue solid line) and at a range of different stellar masses within each of our three environments as indicated. The fraction of SFG within each environment is given in the lower right panel, as a function of environmental over-density $\delta_1$ at redshift $z=0$. The error bars show the $16^{th}$ and $84^{th}$ percentiles of the samples. We also plot the fitted value of sSFR for star forming galaxies within SDSS data for different stellar mass bins as indicated (dotted horizon lines), which are derived from \protect{\cite{Brinchmann2004}} .
            }
    \label{fig:sfr}
\end{figure*}

In figure~\ref{fig:sfr}, we show the relationship between the sSFR of our {\sc Gadget-X} galaxies and the environmental over-density around them at redshift $z=0$.  
We follow \cite{Steinborn2015} in using a specific star formation rate threshold of $sSFR=0.3/t_{Hubble(z)}$ to distinguish between quiescent and star-forming galaxies. Although this definition of a star forming galaxy is somewhat different from the various observational criteria \citep[see.][]{Peng2010, Gabor2012, Muzzin2013}, the precise division line between active and quenched galaxies varies amongst observations. As our simulations are not specifically tuned to match the observed $sSFR-M_*$ relation, we adopt a simple definition for clarity.

In all three of our samples the median sSFR for the all star forming galaxies
(shown as the solid blue line) drops as the over-density, $\delta_1$,
increases. This is principally driven by the changing proportions of galaxies
of different stellar masses. In all cases more massive galaxies have a lower
sSFR and these objects tend to lie preferentially within volumes of greater
local over-density. This transition between a galaxy population at low
over-density that is dominated by small, relatively high sSFR galaxies to one
at high over-density that has a significant population of massive, low sSFR
galaxies leads to the falling trend for the sample as a whole. This same drop
in sSFR is also clearly seen in the observations. 

We claim that the falling sSFR is principally driven by the change with mass and that the relative proportions of galaxies with different masses varies with $\delta$. This is supported by the fact that if we narrow the mass range of galaxies contributing to each of the samples shown in figure~\ref{fig:sfr} the measured sSFR becomes shallower. i.e. for broad mass bands we again see a declining slope in the sSFR - $\delta$ relation.


We also clearly see
(Figure~\ref{fig:sfr}, bottom right panel) that the fraction of star forming
galaxies within each environment is also falling rapidly as the over-density
increases, with very few star forming galaxies above over-densities of 100 in
any of the samples. In the cluster sample this drop in the number of star
forming galaxies appears at lower over-densities, presumably due to the
contamination of this region by back splash galaxies that have passed through
the cluster environment and been quenched due to gas stripping. Another
feature of this panel is that the fraction of star forming
galaxies actually falls at low over-density in both the field and vicinity
galaxy samples. Such low over-densities simply don't exist in the cluster
environment and so no galaxies can possibly reside there. We check components of quenched galaxies in the lowest density bin and find that most of them (over $90\%$) are gas rich. 
There is no mechanism of specially heating the gas in low density region in our simulations. Combining with the fact that the galaxy number in the lowest density bin is ten times less than that in the secondary lowest density bin, we consider the falling a reasonable fluctuation rather than physical phenomenon.

For galaxies within a specified stellar mass range our simulations indicate that the sSFR tends to be rather flat, particularly in the cluster and cluster vicinities samples. In the field the sSFR for the intermediate mass range appears to fall as the over-density increases. The apparent constancy is consistent with previous observations \citep[e.g.][]{Peng2010}. The amplitude of the sSFR is also approximately recovered in all environments.

In Figure~\ref{fig:sfr},  the flat environmental dependencies of median sSFR and the falling fraction of SFG seem to be at odds with each other as the median sSFR of star forming galaxies is not suppressed at higher environmental densities while a higher fraction of galaxies are quenched.  This reproduces the results found by \cite{Peng2010}. \cite{Peng2010} suggested a sharp transition scenario: galaxies continue forming stars at the same rate regardless of their environment, despite the fact that the chance of having been quenched evidently does depend strongly on the environment. This scenario requires instantaneous quenching to occur, and this is the assumption \cite{Peng2010} made in their work.  While such instantaneous quenching might appear idealistic, our simulations support this scenario. Examining star formation histories for individual galaxies we often see sharp drops in SFR during a galaxy's lifetime. Other work suggests that galaxies do not undergo an instantaneous quenching scenario.  Using subhalo abundance matching(SHAM), \cite{Wetzel2013, Muzzin2014} suggested rapid quenching for satellite galaxies with a quenching timescale of less than $0.8 \rm{Gyr}$. \cite{Hahn2017} suggested a longer quenching timescale of $0.5$--$1.5 \rm{Gyr}$ for central galaxies. Currently observational evidence on quenching timescales is lacking, and we therefore cannot confirm which hypothesis truly describes a galaxy's SFR evolution. 

\begin{figure}
    \includegraphics[width=1\linewidth]{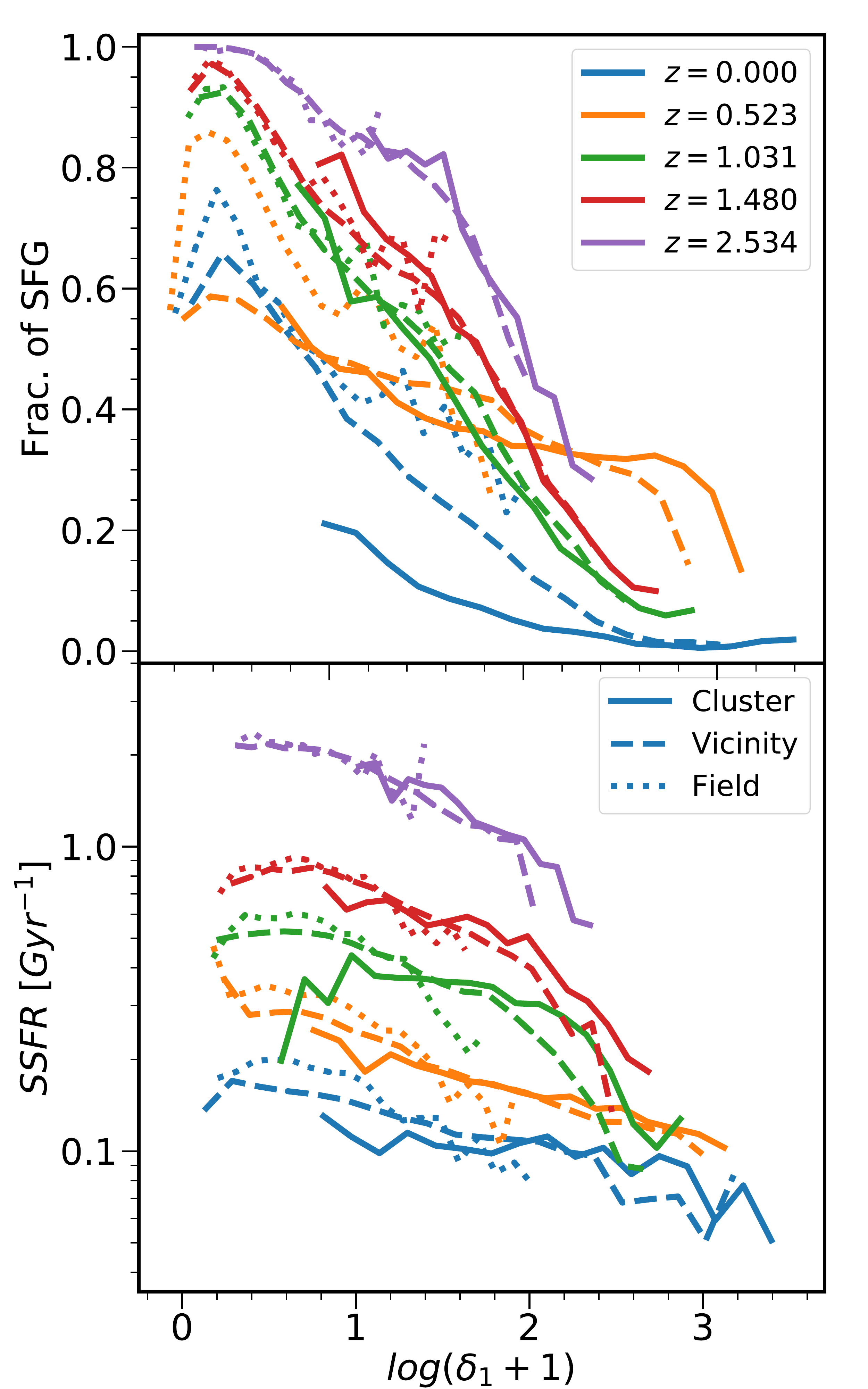}
    \caption{
        Fraction of star forming galaxies (top panel) and median sSFR of star forming galaxy as functions of environmental over-density $\delta_1$ (bottom panel) at various different redshifts as indicated. Galaxies with stellar mass $M_*>10^{9.5}h^{-1}\Msun$ are included. Lines for different redshifts are distinguished with different colours, with different environmental sample groups distinguished via line style. The redshift zero lines are reproduced from Figure~\ref{fig:sfr} for comparison.
            }
    \label{fig:sfr_mz}
\end{figure}

We further investigate how the median sSFR evolves with redshift in Figure~\ref{fig:sfr_mz}, with different line styles indicating our three different samples. All galaxies with stellar mass above $10^{9.5}h^{-1}\Msun$ are included. Figure~\ref{fig:sfr_mz} shows that the fraction of SFG vs. $\delta_1$ (top panel) and $\rm{sSFR}-\delta_1$ (bottom panel) retain a similar shape at all redshifts, with the amplitude decreasing with time. At earlier times three classes of galaxies are in agreement with each other. It is because that at that
time the environments are not obviously distinguishable. 
Essentially, there is a higher fraction of SFG at early times and these galaxies are in general forming stars at a faster rate. There is no sign of any difference between three galaxy classes but there is some evidence that the dependence of sSFR on over-density somewhat steepens at early times.

While our simulations appear to do a reasonable job at reproducing many of the features of the observed sSFR - $\delta_1$ relationship we re-iterate that  our simulations were not tuned to fit this, although many properties are consistent with observations \citep[see.][]{Rasia2015, Biffi2017, Biffi2018a, Truong2018}. One issue is the diversity amongst observational claims.  \cite{Patel2009, Patel2011g} found a declining trend for the sSFR - local density relation of star forming galaxies at $0.6<z<0.9$ and \cite{Muzzin2012} found the sSFR of star forming galaxies increases with the distance from the cluster center at $z\sim1$. These results are consistent with our work. However, there are also works which show no correlation  between these properties at high redshift: \cite{Gruetzbauch2011c, Scoville2013} didn't found any correlation between the sSFR of all galaxies and the local over-density at $z>1$. \cite{Darvish2016} found that the SFR and sSFR of star forming galaxies are not related to the environmental over-density from $z=0.1$ to $z=3$. As \cite{Darvish2016} claimed, the result that no dependence on environment is found at $z\ge1$ may be partly due to the larger photo-z uncertainties at higher redshift and the lack of extremely dense regions in the COSMOS field. Moreover, \cite{Elbaz2007, Cooper2008b, Welikala2016} found an ascending trend of the SFR-density relation at high redshift, which is opposite to what \cite{Patel2009} found or other SFR-density relations at low redshifts. \cite{Popesso2011a} attributed such a reversal in the relation to the contamination of AGN. 

\subsection{Luminosity and Colour Vs. Environmental Over-density}
\label{sec:results:c}

\begin{figure}
    \includegraphics[width=1\linewidth]{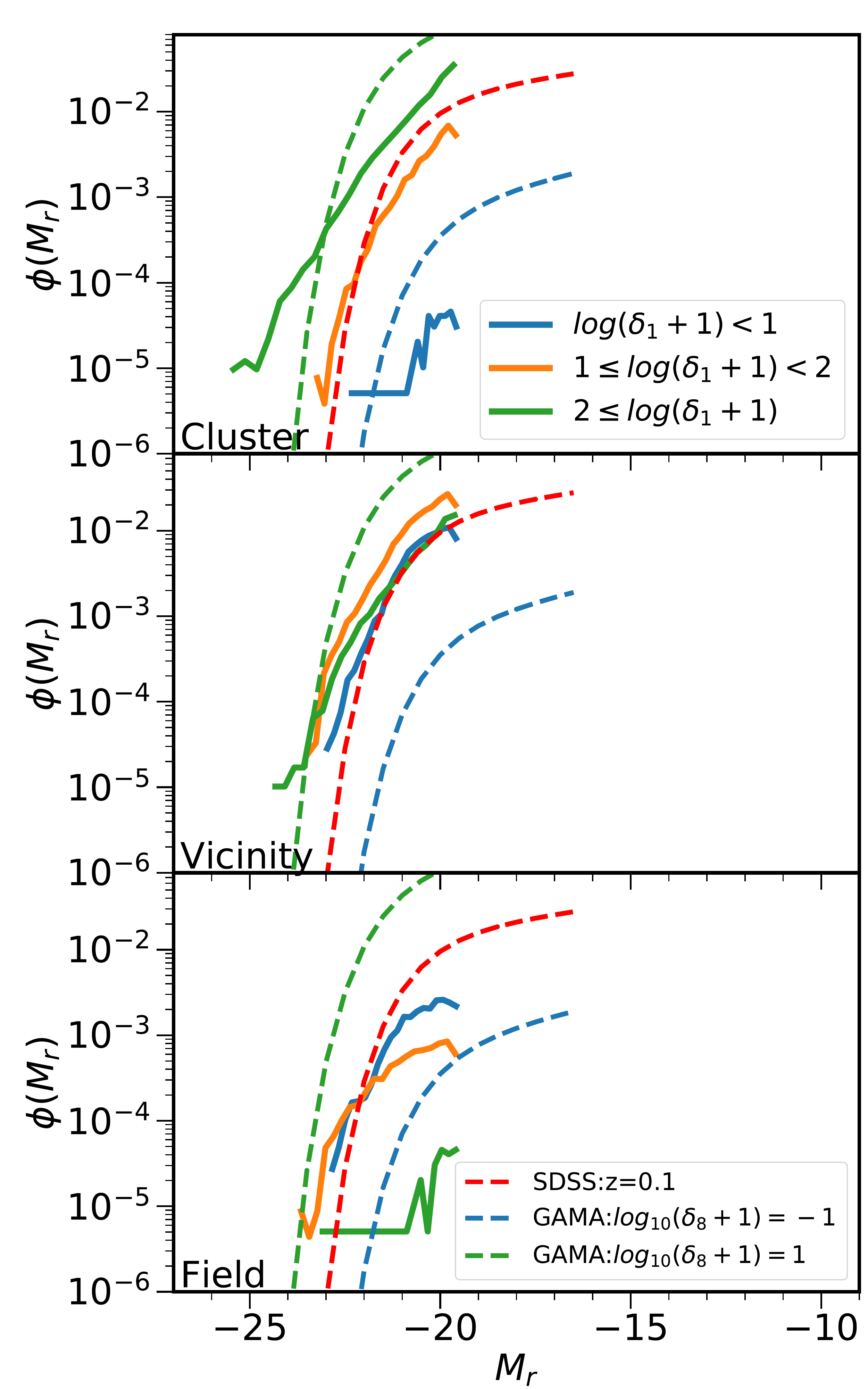}
    \caption{
        The $r$ band luminosity function of galaxies for three bands of environmental over-density for each of our three galaxy samples. As reference, the $^{0.1}r- {\rm band}$ SDSS DR6 luminosity function is shown via the red dashed line. Two best fit lines indicating the luminosity function of under-dense (dashed blue) and over-dense (dashed green) galaxies within the GAMA survey are derived from \protect\cite{McNaught-Roberts2014}.
            }
    \label{fig:lumfunc}
\end{figure}

\begin{figure}
    \includegraphics[width=1\linewidth]{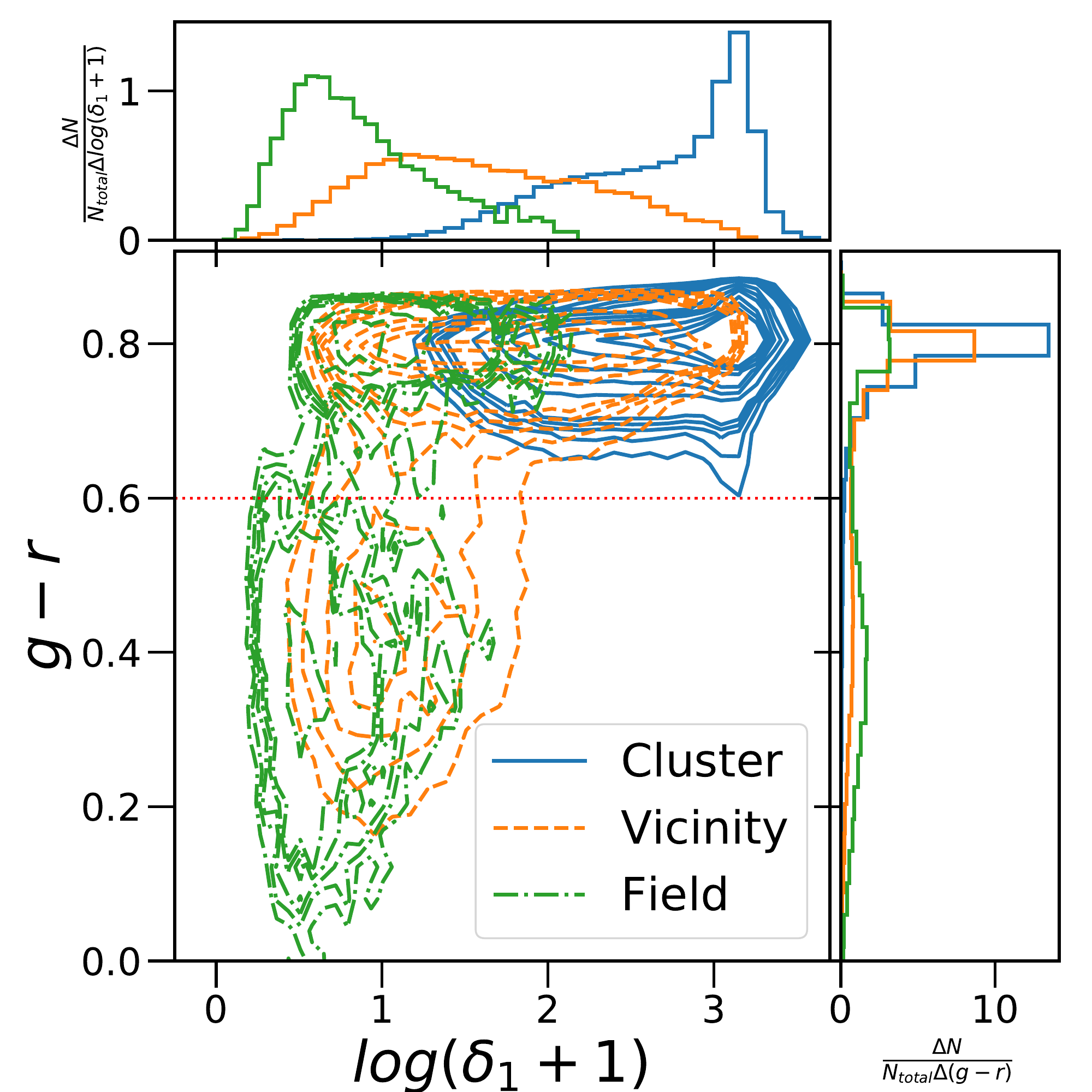}
    \caption{
        In the main panel the contours show the number density of galaxies on the $g-r$ vs. $\delta_1$ plane at redshift $z=0$. Contours for each of our three sample groups are distinguished by different line colours and styles. The horizontal red dotted line at $g-r=0.6$ indicates the rough division between red and blue galaxies at redshift $z=0$. In the top and right sub plots, the histograms show the galaxy number density in each over-density/color bin again split by galaxy sample.
            }
    \label{fig:color}
\end{figure}

\begin{figure*}
    \includegraphics[width=1\linewidth]{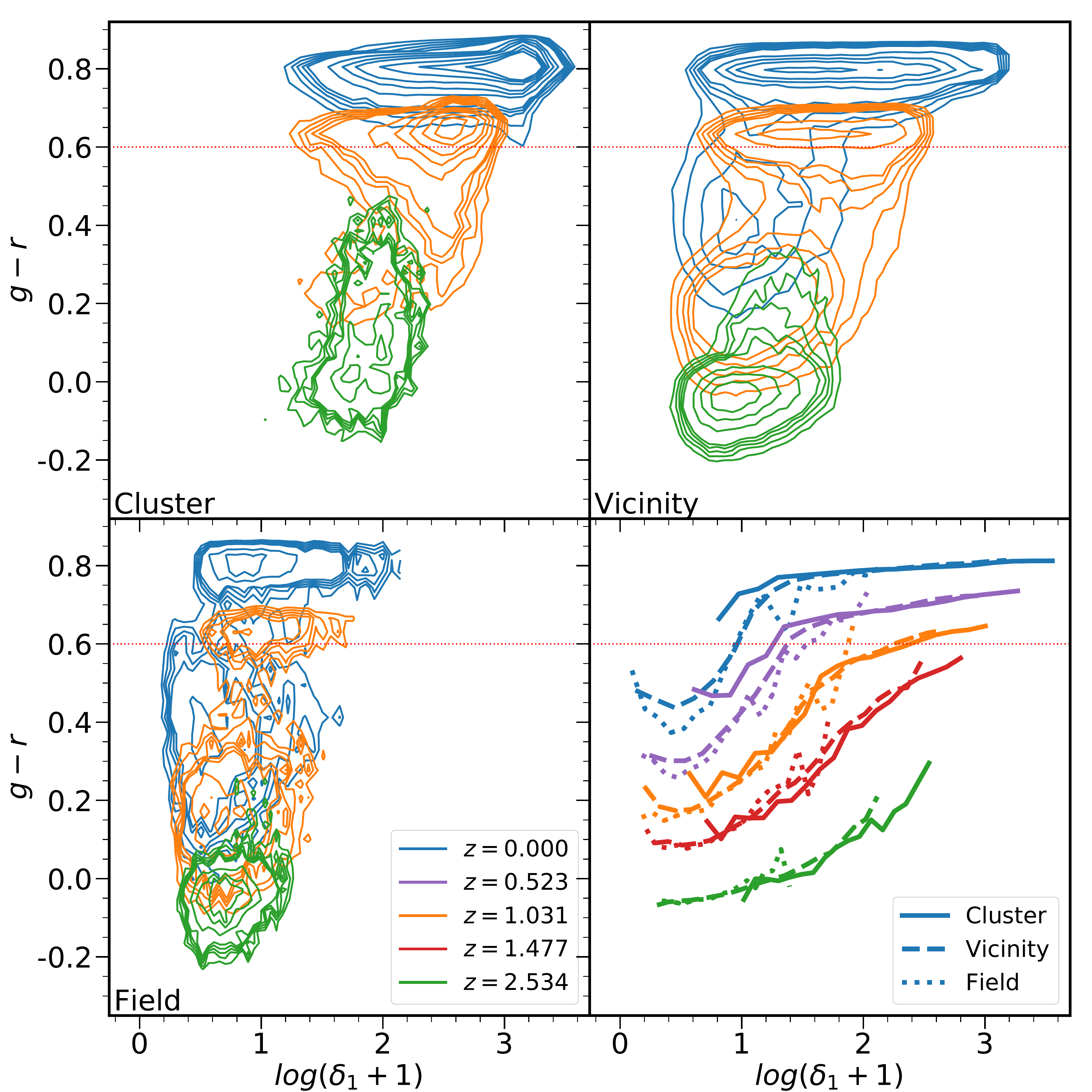}
    \caption{
        Contours of galaxy number density on the $g-r$ vs. $\delta_1$ plane
        at different redshifts as indicated by the various colours. Three of
        the panels present the results from each of our three sample groups
        as indicated. For brevity, only contours at $z=0$, $z=1.031$ and $z=2.534$ are shown. The bottom right hand panel presents the median
        relationship between $g-r$ and $\delta_1$ split via line style for
        each sample group and by colour as a function of redshift. The
        horizontal red dotted line at $g-r=0.6$ indicates the rough division
        between red and blue galaxies at redshift $z=0$.
            }
    \label{fig:color_mz}
\end{figure*}

First we present the $r$ band luminosity function for three bands of environmental over-density, $\delta_1$ for each of our three different sample groups in Figure~\ref{fig:lumfunc}.
The Luminosity function for galaxies within a specific density bin is measured following the method described in Section 2.5 of \citep{McNaught-Roberts2014}
The dashed lines indicate best fits to the SDSS luminosity function (red) and galaxies within under- (blue) and over- (green) dense environments within the GAMA survey. The green solid lines in each panel indicate galaxies with the highest local over-density, $\delta_1$. There are lots of these in the cluster sample and they are generally brighter and more massive than the galaxies in less over-dense environments. Even though we have removed the central galaxies from each of the 324 clusters (these are hard to distinguish from their surrounding intra-cluster light) there remain some extremely bright objects. This indicates that the treatment of large central galaxies within the {\sc Gadget-X} model could be improved. The excess here is due to central galaxies within large infalling structures. This effect for the most over-dense galaxies persists into the vicinities (middle panel) and occurs for the same reason: there are some large groups with dominant central galaxies here. In the field sample there are no large groups by construction and so there are very few galaxies in the most over-dense environment. For the orange and blue solid lines indicating galaxies in intermediate and low over-density local environments the luminosity function in all three of our sample regions is not unrealistic when compared to the observational best fits. The shape is well recovered in all cases with, as expected, the amplitude of the under-dense luminosity function rising as we move from the cluster to vicinities to field samples. The observed luminosity function is built up from galaxies within all three of our sample regions and is a combination of them with unknown weights. Essentially, unless a very large unbiased survey has been constructed the precise amplitude of the luminosity function curves will depend strongly on the sample selection.

Secondly, we check the colour-$\delta_1$ relation in  Figure~\ref{fig:color}. We use the difference between the magnitude in the $g$ and $r$ bands, $M_g - M_r$ to evaluate the colour of each galaxy and display this as a function of over-density, $\delta_1$. Contours of different colours indicate the galaxy number density at $z=0$ within each of our three sample regions as indicated in the legend. The horizontal dotted red line at $g-r=0.6$ indicates the rough division between red and blue galaxies at redshift $z=0$. Clearly, within the cluster sample the vast majority of the galaxies are red at $z=0$, with almost no blue galaxies. There are also no galaxies with low environmental over-density because such an environment doesn't exist for this sample. The field sample contains examples of both red and blue galaxies and, as previously seen, does not contain any high over-density environment (and so has no high over-density galaxies). The cluster vicinities sample contains galaxies with both colours and environments, although it does not probe either the highest or the lowest over-densities. The clear bi-modality between blue and red galaxies in the observed galactic population is well recovered by the {\sc Gadget-X} simulations.

Finally we examine the evolution of the colour-$\delta_1$ plane in Figure~\ref{fig:color_mz}. Three panels show via coloured contours the change in the distribution of galaxy number density on the $g-r$ vs. $\delta_1$ plane as a function of redshift. Each panel displays results from a different sample region as indicated. As before, the red dotted horizontal line at $g-r=0.6$ indicates the rough division between red and blue galaxies at redshift $z=0$. Clear evolution in the colour of the galactic population in all three samples is seen and this mirrors the expected trend. In all three samples an obvious red sequence is visible and in-place at $z=1$ and colour bi-modality is also evident. At $z \sim 2.5$ the majority of galaxies in all environments are blue, even if the evolution of the dividing line between red and blue galaxies is taken to evolve with redshift. The bottom right panel displays the evolution of the median values within each sample (distinguished via line type) as a function of redshift (distinguished via line colour). The median colour as a function of over-density within all three of our samples is largely indistinguishable, indicating that to a large extent the apparent differences in colour as a function of $\delta$ are largely driven by how well the over-density range is spanned by the sample rather than via any difference in the galaxy's colour due to its environment.

\subsection{Local Over-density Vs. Halo Mass}
One critical issue is what environmental measure is most closely related to a galaxy's properties. Based on the theory of hierarchical structure formation, it is quite natural to think that a galaxy's properties should be closely linked to their host virialized dark matter halo. On the other hand, many observations found a close relationship between local over-density and galactic properties. Our simulations support the observational result, as we find that local over-density is a more direct measure of a galaxy's properties than its host halo mass. 

An intuitive way to quantify the relationship between two variables is by  calculating Spearman's rank correlation coefficient. Spearman's rank correlation coefficient $r_s$ is defined as:
\begin{equation}
r_s = \frac{cov(rg_X, rg_Y)}{\sigma_{rg_X}\sigma_{rg_Y}}
\end{equation}
Here $rg_{Xi}$ is the rank of $i^{th}$ element in the array of variable $X$. The value of $r_s$ indicates how well the relationship between two variables can be described using a monotonic function. The range of $r_s$ is $[-1,1]$, $r_s=0$ indicating no relationship between the two variables. $r_s=1$ indicates a perfect positive relationship, while $r_s=-1$ indicates a perfect negative relationship. The larger the value of $|r_s|$, the better the relationship is. 

\begin{table}
	\begin{center}
    \caption{Spearman's correlation coefficient $r_s$ for the relationship between galaxy properties and local over-density, halo mass and stellar mass. $r_s$ for our cluster, vicinity and field galaxy samples are listed in separate tables.}
    \label{tab:r}
	\begin{tabular}{c| c c c}
    \toprule
    Cluster \\
    \toprule
	$r_s$ & $\rm{Frac_{SFG}}$ & $\rm{sSFR_{SFG}}$ & $g-r$ \\ 
	\midrule
    $\delta_1$ & -0.932 & -0.099 & 0.311\\
    $M_{halo}$ & 0.809 & -0.280 & -0.104\\
    $M_*$ & 0.924 & -0.321 & 0.388\\
    \bottomrule
    Vicinity \\
    \toprule
	$r_s$ & $\rm{Frac_{SFG}}$ & $\rm{sSFR_{SFG}}$ & $g-r$ \\ 
	\midrule
    $\delta_1$ & -0.962 & -0.161 & 0.454\\
    $M_{halo}$ & 0.528 & -0.041 & 0.009\\
    $M_*$ & -0.364 & -0.209 & 0.398\\
    \bottomrule
    Field \\
    \toprule
	$r_s$ & $\rm{Frac_{SFG}}$ & $\rm{sSFR_{SFG}}$ & $g-r$ \\ 
	\midrule
    $\delta_1$ & -0.946 & -0.178 & 0.373\\
    $M_{halo}$ & 0.502 & -0.111 & -0.293\\
    $M_*$ & -0.490 & -0.250 & 0.421\\
    \bottomrule
    \end{tabular}
	\end{center}
\end{table}

\begin{table*}
	\begin{center}
    \caption{The same as Table \ref{tab:r}, but the samples are divided into three stellar mass bins.}
    \label{tab:r2}
	\begin{tabular}{c | c c c | c c c | c c c}
    \toprule
    Cluster & \multicolumn{3}{|c|}{$M_*<10^{10}h^{-1}\Msun$} & \multicolumn{3}{|c|}{$10^{10}h^{-1}\Msun<M_*<10^{11.5}h^{-1}\Msun$} & \multicolumn{3}{|c}{$M_*>10^{11.5}h^{-1}\Msun$} \\
    \toprule
	$r_s$ & $\rm{Frac_{SFG}}$ & $\rm{sSFR_{SFG}}$ & $g-r$ & $\rm{Frac_{SFG}}$ & $\rm{sSFR_{SFG}}$ & $g-r$ & $\rm{Frac_{SFG}}$ & $\rm{sSFR_{SFG}}$ & $g-r$ \\ 
	\midrule
    $\delta_1$ &  -0.883 &  0.082 &  0.446 & -0.933 &  0.076 &  0.326 & 0.939 & -0.013 & -0.400  \\
    $M_{halo}$ &  0.345 &  0.085 & -0.718 & 0.285 & -0.162 & -0.304 & 0.867 & -0.172 & -0.738 \\
    $M_*$      & 0.818 & -0.076 &  0.283 & 0.552 & -0.143 &  0.102 &  0.564 & -0.095 & -0.731 \\
    \bottomrule
    Vicinity & \multicolumn{3}{|c|}{$M_*<10^{10}h^{-1}\Msun$} & \multicolumn{3}{|c|}{$10^{10}h^{-1}\Msun<M_*<10^{11.5}h^{-1}\Msun$} & \multicolumn{3}{|c}{$M_*>10^{11.5}h^{-1}\Msun$} \\
    \toprule
	$r_s$ & $\rm{Frac_{SFG}}$ & $\rm{sSFR_{SFG}}$ & $g-r$ & $\rm{Frac_{SFG}}$ & $\rm{sSFR_{SFG}}$ & $g-r$ & $\rm{Frac_{SFG}}$ & $\rm{sSFR_{SFG}}$ & $g-r$ \\ 
	\midrule
    $\delta_1$ & -0.964 & -0.080 &  0.638 & -1.0 & -0.171 & 0.396 & 0.127 & -0.026 & -0.456   \\
    $M_{halo}$ &  0.818 &  0.134 & -0.750 & 0.152 & -0.025 & -0.088 & 0.033 & -0.106 & -0.458 \\
    $M_*$      &  - & -0.082 &  0.013 & -0.685 & -0.199 & 0.169 & 0.762 & 0.033 & -0.499 \\
    \bottomrule
    Field & \multicolumn{3}{|c|}{$M_*<10^{10}h^{-1}\Msun$} & \multicolumn{3}{|c|}{$10^{10}h^{-1}\Msun<M_*<10^{11.5}h^{-1}\Msun$} & \multicolumn{3}{|c}{$M_*>10^{11.5}h^{-1}\Msun$} \\
    \toprule
	$r_s$ & $\rm{Frac_{SFG}}$ & $\rm{sSFR_{SFG}}$ & $g-r$ & $\rm{Frac_{SFG}}$ & $\rm{sSFR_{SFG}}$ & $g-r$ & $\rm{Frac_{SFG}}$ & $\rm{sSFR_{SFG}}$ & $g-r$ \\ 
	\midrule
    $\delta_1$ & -0.927 & -0.078 &  0.698 & -0.988 & -0.216 & 0.310 & - & -0.335 & 0.335 \\
    $M_{halo}$ & 0.964 & -0.132 & -0.811 & 0.133 & -0.129 & 0.270 & - & -0.415 & 0.340 \\
    $M_*$      &  - & -0.082 &  0.136 & -0.358 & -0.317 & 0.266 & - & -0.400 & 0.137 \\
    \bottomrule
    \end{tabular}
	\end{center}
\end{table*}
In Table~\ref{tab:r}, we tabulate the value of $r_s$ between galaxy properties and three environmental measures, namely $\delta_1$, the host halo mass and the galaxy stellar mass. We calculate this measure for all three of our galaxy samples, as indicated in the table. The value of $r_s(\delta_1)$ is quite consistent among our three samples. As we have already seen above: the fraction of SFG is strongly negatively correlated with $\delta_1$; the sSFR of SFG is slightly negatively correlated with $\delta_1$; the colour, $g-r$, is positively correlated with $\delta_1$. The $r_s(M_{halo})$ shows a quite different relationship, as well as changing significantly across the three galaxy samples. This variation in $r_s(M_{halo})$ implies that the influence of host halo differs in different large scale environments. Thus halo mass is a worse indicator of local environment than local over-density $\delta_1$. This should not necessarily be a surprise, as a large galaxy cluster contains many thousands of galaxies with a wide variation in properties and so the single halo mass should not be expected to correlate particularly well on this level.

For comparison we also calculate Spearman's correlation coefficient for galaxy properties and stellar mass. In this case, for the cluster sample, $r_s$ has a value of $0.924$ for the fraction of star forming galaxies, which indicates a significant positive relationship. This relationship is opposite to that observed due to the essential lack of low-mass SFG in the cluster sample. Except for this, the values of the other $r_s(M_*)$ meet our expectations well. By comparing $r_s(M_*)$ and $r_s(\delta_1)$, we find that the fraction of SFG is more influenced by the local over-density, while the sSFR of SFG is more influenced by the stellar mass. For galaxy color, it seems that both $\delta_1$ and $M_*$ have similar weights.

The accuracy of $r_s$ could suffer due to the incompleteness of 
our samples. Thus we divided our galaxies into different stellar mass bins to test the 
sensitivity of $r_s$ to our galaxy sample. As Table \ref{tab:r2} shows, the $r_s$ for the low mass and high mass ends changes a lot compared to Table \ref{tab:r}, while the $r_s$ for the median stellar mass range keeps the same trends and similar
values. Except for massive galaxies, the strong influence of environment, $\delta_1$, on the fraction of SFG is stable, which further prove the important link between environment and the SFG fraction. The negative correlation 
between $M_*$ and sSFR of SFG is still clear, although it becomes weaker in the low and high stellar mass range. The $r_s(M_{halo})$ becomes variable 
when the different stellar mass ranges are applied. The $r_s$ between color $g-r$ 
and $\delta_1$, $M_*$, $M_{halo}$ is very similar to Table \ref{tab:r}, except 
for massive galaxies. $r_s(M_{halo}, g-r)$ shows negative relation in 
most places, which is again counter intuitive. In summary, we consider that the influence of $\delta_1$
is much less sensitive to mass completeness than $M_{halo}$.

\section{Discussion \& Conclusions} \label{sec:summary}

In this work, we have examined the local environmental dependence, measured via the over-density relative to the cosmic mean on a scale of $1\Mpc/h$, $\delta_1$, of several galactic properties. We do this with three galaxy samples which broadly delineate the large scale environment, specifically a field sample constructed from four field re-simulations of volumes that do not contain any large halo at $z=0$, a cluster sample that contains all galaxies within twice $R_{200}$ of the centre of the largest halo within each of our 324 cluster re-simulations and a `vicinity' sample that consists of all those galaxies not in the cluster sample but within the uncontaminated region of the cluster re-simulations. These latter vicinity galaxies are by definition relatively close to a large structure. Cluster galaxies are essentially within a large structure (the median mass of the our cluster halos is $8 \times 10^{14}\Msun/h$). Field galaxies are a long way from any major dark matter halo, with the largest halo contained within them having a mass of $4 \times 10^{13}\Msun/h$. 

One primary advantage of our project ``{\sc The Three Hundred}'' \citep{Cui2018} is the
abundance of simulated galaxies. Since we in total employed 328
re-simulations, the number of available galaxies at any one time exceed
$200,000$, which is almost at the same level as observations, i.e.,
\cite{Peng2015} select $238,474$ redshift reliable galaxies from SDSS DR7 for
analysis. Such an abundance of galaxies allow us to test the dependence of
galaxy properties on environment convincingly, as well as other relations.
In addition, we employ an advanced hydro dynamical code {\sc Gadget-X} which reasonably recover the observed galactic SFR-$M_*$ relation.

There are three main environmental relations explored in this work:

\begin{itemize}
\item {First, the environmental over-density $\delta_1$ has no influence on the measured specific star formation rate of star forming galaxies. 
Although, the sSFR for all galaxies {\it does} show a trend with over-density, this is driven by the changing balance of galactic mass with environment. 
At fixed stellar mass, the sSFR is independent of the local over-density. 
There is a larger fraction of massive galaxies in over-dense environments than in under-dense environments, and the sSFR is clearly dependent on the mean stellar mass of the sample. This mirrors the observational result found by SDSS, where the sSFR also drops as the median galaxy mass increases.}

\item {Second, as expected, the fraction of star forming galaxies declines as the over-density increases in all three of our samples. Our three classes of galaxies have different SFG Fraction-$\delta_1$ relation at $z=0$, but all three curves converge at high redshift. This implies that, in spite of local over-density, the large scale structure, such as the presence of a cluster or a void, affects the quenching process. At high redshift, these large scale structures have not formed and thus their influence does not yet show up.}

\item Third, galaxies become redder in higher over-density environments. The colour distribution of galaxies with over-density is bimodal as expected, with a clear red sequence and blue cloud. 
Median colours as a function of over-density compared in the field, vicinity and cluster samples do not change. This is somewhat surprising given the visual impression of the contour plots. What is actually happening is that different regions of the over-density axis are being probed but the underlying distribution is largely unchanged.
\end{itemize}

We further qualified three environmental relations above with Spearman's correlation coefficient $r_s$. The value of $r_s$ for each galaxy property-$\delta_1$ relation is in agreement with what we have seen in the figures. 
Spearman's correlation coefficient $r_s$ for galaxy properties and both $\delta_1$ and $M_{halo}$ is also compared. We found that $r_s$ for the relationship between $\delta_1$ and galaxy properties is much more stable than the relationship with halo mass. This implies that, compared with halo mass, local over-density is more likely to be directly linked to a galaxy's properties.
We stress that Spearman's correlation coefficient $r_s$ is a parameter evaluating the monotonic relation. It is possible that the relationship between a galaxy's properties and host halo mass is not monotonic.

{Our galaxy sample recovers many observational properties well. But it is not yet perfect. Within the set of {\sc Gadget-X} galaxies some are unrealistically bright. These are the dominant central objects within large dark matter halos. Future iterations of the {\sc Gadget-X} code will examine this issue further. This aside, the luminosity function of the galaxies produced by {\sc Gadget-X} has the correct shape. The amplitude is dependent on the relative abundance of that particular environment within the sample and so is largely driven by sample selection. Care should be taken when constructing galaxy luminosity functions to avoid this dependence on local environment.}

To summarize, the dependence of galaxy star formation rates and colour on
over-density for field, vicinity and cluster samples is examined as a
function of redshift using a large sample of 324 massive galaxy clusters. We
show that the {\sc Gadget-X} galaxies within these objects (and also in field
regions) generally form a reliable sample which can be used to test
observational inferences upon the formation and evolution of the galactic
population.

\acknowledgments
{\bf Acknowledgment:}
This work has received financial support from the European Union's Horizon
2020 Research and Innovation programme under the Marie Sklodowskaw-Curie
grant agreement number 734374, i.e. the LACEGAL project.

The authors would like to thank The Red Espa\~nola de Supercomputaci\'on for
granting us computing time at the MareNostrum Supercomputer of the BSC-CNS
where most of the cluster simulations have been performed. Part of the
computations with Gadget-X have also been performed at the
'Leibniz-Rechenzentrum' with CPU time assigned to the Project 'pr83li'. The
authors would like the acknowledge the Centre for High Performance Computing
in Rosebank, Cape Town for financial support and for hosting the ``Comparison
Cape Town" workshop in July 2016. The authors would further like to
acknowledge the support of the International Centre for Radio Astronomy
Research (ICRAR) node at the University of Western Australia (UWA) in the
hosting the precursor workshop ``Perth Simulated Cluster Comparison" workshop
in March 2015; the financial support of the UWA Research Collaboration Award
2014 and 2015 schemes; the financial support of the ARC Centre of Excellence
for All Sky Astrophysics (CAASTRO) CE110001020; and ARC Discovery Projects
DP130100117 and DP140100198. We would also like to thank the Instituto de
Fisica Teorica (IFT-UAM/CSIC in Madrid) for its support, via the Centro de
Excelencia Severo Ochoa Program under Grant No. SEV-2012-0249, during the
three week workshop ``nIFTy Cosmology" in 2014, where the foundation for this
this project was established.

YW is supported by the NSFC No.11643005. LLF is supported by the NSFC
No.11733010. WC, AK and GY are supported by the {\it Ministerio de
Econom\'ia y Competitividad} and the {\it Fondo Europeo de Desarrollo
Regional} (MINECO/FEDER, UE) in Spain through grant AYA2015-63810-P. AK is
further supported by the Spanish Red Consolider MultiDark FPA2017‐90566‐REDC
and also thanks Nohelani Cypriano for lihue.

YW thanks Weishan Zhu and Jiaxin Han for their useful comments and
suggestions. We also thanks Giuseppe Murante for the help with running the {\sc Gadget-X} simulations.

The authors contributed to this paper in the following ways: GY, FRP, AK, CP
and WC formed the core team that provided and organized the simulations and
general analysis of data. 
GY and SG ran at LRZ Munich the underlying MDPL2 simulation and selected the regions for the re-simulations. 
AA \& Giuseppe Murante managed the {\sc Gadget-X} simulation and
provided the data. The specific data analysis for this paper was led by YW.
YW, FRP and AK wrote the text. All authors had the opportunity to
proof read and provide comments on the paper.

\bibliographystyle{aasjournal}
\bibliography{Lib}
\label{lastpage}
\end{document}